\documentclass[pra, showpacs, twocolumn, floatfix]{revtex4}
\usepackage{graphicx}
\usepackage{amsmath, amsfonts, amssymb, bm}
\usepackage[]{psfrag}
\psfragscanoff 
\begin{document}
\title{Quantum tunneling through vacuum-multiparticle induced potentials}
\author{Mihai A. \surname{Macovei}}
\email{mihai.macovei@mpi-hd.mpg.de}
\affiliation{Max-Planck Institute for Nuclear Physics, Saupfercheckweg 1,D-69117 Heidelberg, Germany}
\date{\today}
\begin{abstract}
The vacuum cavity mode induces a potential barrier and a well when an ultra-slow excited atom enters the 
interaction region so that it can be reflected or transmitted with a certain probability. We demonstrate 
here that a slow-velocity excited particle tunnels freely through a vacuum electromagnetic field mode filled 
with $N-1$ ground state atoms. The reason for this is the trapping of the moving atom into its upper 
state due to multiparticle influences and the corresponding decoupling from the interaction with the 
environment such that the emitter does not {\it see} the induced potentials.
\end{abstract}
\pacs{32.80.-t, 42.50.Pq, 42.50.Fx, 42.65.Tg}
\maketitle
The technological progress in the area of the Cavity Quantum Electrodynamics made possible to achieve
the strong-coupling regime where the coherent interaction between a single atom and the light field
dominates over the dissipation processes \cite{CuD,AG1,exp1,exp2,HW,RBH,NS,WM,SZ,OL,RT,ST}. Experiments 
showing vacuum  Rabi splitting \cite{AG1,exp1}, ringing regime of superradiance \cite{exp2}, collapses 
and revivals as well as the trapping states in the micromaser \cite {HW}, and the quantum entanglement 
\cite{HW,RBH} have been already reported. In addition, creation of atomic nanostructures \cite{NS}, atomic 
interferometry and diffraction by a light wave \cite{WM,SZ} or a laser oscillation with one atom in an 
optical resonator \cite{OL} were successfully demonstrated. Another significant experimental result in 
the strong coupling domain was the real-time detection of single atoms transiting through high-finesse 
optical cavities \cite{RT}. Interesting phenomena are expected from the super-strong coupling regime \cite{ST}.

Remarkably, ground-state atoms accelerated through a vacuum - state cavity radiate real photons \cite{acc},
while particle acceleration by stimulated emission of radiation has been experimentally demonstrated in \cite{paser}.
On the other hand, the atomic motion of single ultraslow particles passing through high-quality resonators can 
be considerable modified due to the fact that the kinetic energy of the atoms is comparable with their coupling 
energy to the cavity field \cite{USL,AAT,exp3}. The tunneling of matter wave packets through different types of
traps and barriers attracted then a lot of attention due to a possible design of an atomic soliton laser \cite{SL} 
or a velocity selector for ultracold particles \cite{VSL}. 

Here we report on the resonant quantum tunneling of slow atoms through vacuum-multiparticle induced 
potentials in high-finesse resonators. An atomic wave packet incident on a vacuum cavity mode which 
induces a potential barrier and a well can be reflected or transmitted with a certain probability. 
The magnitudes of these potentials however are significantly enhanced when in the interaction region 
there are many indistinguishable ground-state particles. The tunneling mechanism relies then on the 
trapping of the moving atom into its upper state due to the multiparticle interactions and the 
corresponding decoupling from the environment such that the emitter does not {\it see} the induced 
potentials. Larger atomic systems lead to a complete transfer of single excited 
particles through the vacuum - multiparticle induced potentials. The effect is independent on the width 
of these potentials and on particle's velocities providing that its kinetic energy is in the classically 
forbidden range. For moderate samples the transmission probability is characterized by sharp maxima and 
minima which can be useful in determining the number of the trapped ground-state particles inside the 
resonator within one emission wavelength.

We consider a monoenergetic beam of excited two-level slow atoms, each of mass $\mu$, to be incident 
upon a single-mode high-quality cavity of length $L$ filled with $N-1$ identical ground-state motionless 
particles. The atomic flux is adjusted so that only one particle interacts with the cavity vacuum field 
at a time. The dipole-dipole interactions between the radiators are neglected here assuming lower atomic 
densities. The Hamiltonian describing such a multiatom sample resonantly interacting with the quantized 
cavity mode in the dipole and rotating-wave approximation is given by:

\begin{eqnarray}
H &=& \hbar\omega_{c}a^{\dagger}a + \sum_{l \in \{1,\cdots,N \}}\hbar\omega_{0l}R_{zl} 
+ \frac{p^{2}_{z}}{2\mu} \nonumber \\
&+& \sum_{l \in \{1,\cdots,N\}}\hbar\bigl (g_{l} R^{-}_{l}a^{\dagger} + g^{\ast}_{l}a R^{+}_{l}\bigr ). \label{Hm}
\end{eqnarray}

Here $a^{\dagger}$ and  $a$ are the radiation creation and annihilation operators obeying the commutation relations
$[a,a^{\dagger}]=1$, and $[a,a]=[a^{\dagger},a^{\dagger}]=0$. The atomic operator $R^{+}_{l}(R^{-}_{l})$ describes 
the excitation (de-excitation) of $l$th atom into its upper (lower) energy level and satisfies the usual commutation 
relations for su(2) algebra, i.e $[R^{+}_{j},R^{-}_{l}]=2R_{zj}\delta_{jl}$ and $[R_{zj},R^{\pm}_{l}]=\pm 
R^{\pm}_{j}\delta_{jl}$ with $R_{z}$ being the inversion operator. 

In Eq.~(\ref{Hm}) the first and the second terms are the free electromagnetic field (EMF) and free atomic Hamiltonians, 
respectively. The third term is the kinetic energy of the center-of-mass motion (CM) of the atom probing the compound 
system, i.e. 'cavity plus ground state atoms', while $p_{z}=\hbar \chi$ is the quantum-mechanical motional momentum 
operator of this particular emitter. The fourth term takes into account the interaction of all atoms with the vacuum 
cavity field. Further, $g_{l}$ are the atom-field couplings strength for the interaction between the quantized field 
with frequency $\omega_{c}$ and the atoms with level spacing $\hbar\omega_{0l}$. Note that the motion of the ground-state 
particles is neglected here assuming fixed positions for them. Though a ground state atom recoils with a momentum $\hbar k_{f}$ 
when it absorbs a photon of wave vector $k_{f}$ this process is less probable here since, as it will be shown, there is no
photon emissions in the multiparticle case. Moreover, thermal fluctuations will not affect the ground-state particles if 
the mean thermal photon number $\bar n$ inside the resonator mode is very small, i.e. $\bar n \to 0$ \cite{mek}.
\begin{figure}[t]
\includegraphics[height=3.8cm]{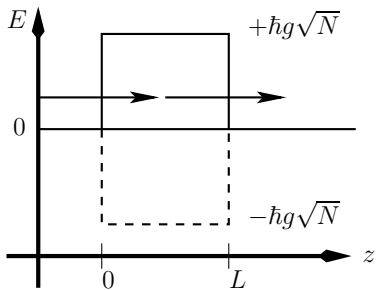}
\caption{\label{fig-1} An excited atom with CM kinetic energy $E=(\hbar \chi)^{2}/2\mu$ (indicated by two horizontal
arrows) is incident from the left on a cavity of length $L$ that contains $N-1$ ground state particles. The effect of 
the cavity with a z-independent mode function is to induce a rectangular potential barrier ($+\hbar g\sqrt{N}$) and 
a rectangular well ($-\hbar g\sqrt{N}$) while the excited atom tunnels through it if $N \gg 1$.}
\end{figure}

%
As there is only one excitation at a time, the atom-cavity system can be investigated in the following basis states: 
$|1\rangle=|e_{1},g_{2},\cdots,g_{N};0\rangle$, $|j\rangle=|g_{1},g_{2},\cdots,e_{j},\cdots,g_{N};0\rangle$ ($j=2,\cdots,N$) 
and $|0\rangle=|g_{1},g_{2},\cdots,g_{N};1\rangle$. Here $|1\rangle$ means that the particle which enters 
the cavity is excited while all other atoms are in their ground states inside the vacuum-state resonator. 
Respectively, the state $|j\rangle$ ($j\not=1$) is a collective state of the $j$th atom being excited 
into the vacuum cavity mode. All the atoms in their ground states and one photon into the cavity mode 
is described by $|0\rangle$. Due to the equivalence of $N-1$ ground state motionless atoms with respect 
to the absorption-emission processes of a single cavity photon the collective states $|j\rangle$ are 
identical simplifying considerable the analytical treatment of the problem. That is to say the atoms passing through
the interaction region and those $N^{'} = N-1$ atoms inside the resonator are distinguishable. In this case $|j\rangle$ 
can be treated as a coherent atomic state \cite{mek} which denotes a symmetrized $N^{'}-$atom state in which $N^{'}-1$ atoms are 
in their ground levels while one atom being excited into its upper bare state, respectively. Then the atom-field interaction 
Hamiltonian in Eq.~(\ref{Hm}), i.e. 
\begin{eqnarray*}
H_{af}=\sum_{l}\hbar\bigl (g_{l} R^{-}_{l}a^{\dagger} + g^{\ast}_{l}a R^{+}_{l}\bigr ),  
\end{eqnarray*}
can be diagonalized using the above mentioned basis states to obtain the following expressions for the normalized eigenfunctions
\begin{eqnarray}
|\Psi^{(+)}_{N}\rangle &=& \frac{1}{\sqrt{2N}}\bigl \{|1\rangle + \sqrt{N-1}|j\rangle \bigr \} 
+ \frac{1}{\sqrt{2}}|0\rangle, \nonumber \\
|\Psi^{(0)}_{N}\rangle &=& \frac{1}{\sqrt{N}}\bigl \{\sqrt{N-1}|1\rangle - |j\rangle \bigr \}, \nonumber \\ 
|\Psi^{(-)}_{N}\rangle &=& \frac{1}{\sqrt{2N}}\bigl \{|1\rangle + \sqrt{N-1}|j\rangle \bigr \} 
- \frac{1}{\sqrt{2}}|0\rangle, \label{fp}
\end{eqnarray}
with eigenvalues $\lambda^{(+)}=\hbar g\sqrt{N}$, $\lambda^{(0)}=0$ and $\lambda^{(-)} = -\hbar g\sqrt{N}$, respectively. 
Note that we have considered here the exact resonance between the cavity mode frequency and atomic ones, i.e. 
$\omega_{c}=\omega_{0} \equiv \omega_{01}=\omega_{02}=\cdots=\omega_{0N}$. In addition, equal atom-field 
real couplings ($g_{1}=g_{2}=\cdots g_{N} \equiv g$) were assumed while the field mode in the cavity is z-independent, 
i.e. as for an idealized mesa function shape (see Fig.~\ref{fig-1}). Thus, the collective behaviors we consider are 
entirely due to the mutual coupling of all the emitters with a common radiation field. 

We assume the initial wave packet of the moving particle to be $\phi(z,t)=\int d\chi A(\chi)e^{-i(\hbar \chi^{2}/2\mu)t}e^{i\chi z}$, 
where $A(\chi)$ is a normalized function of $\chi$ (say, for instance, a Gaussian) peaked about the mean momentum $\hbar \bar \chi$.
The combined state of the atom-cavity system at the initial time $t=0$ is $|\Phi(z,0)\rangle=\phi(z,0)|1 \rangle$, 
where $|1\rangle$ has to be represented via dressed state functions, that is 
\begin{eqnarray*}
|1\rangle = [|\Psi^{(+)}_{N}\rangle + \sqrt{2(N-1)}|\Psi^{(0)}_{N}\rangle + |\Psi^{(-)}_{N}\rangle]/\sqrt{2N}. 
\end{eqnarray*}
Now the initial state will be expressed as a sum of components of the
form $|\Phi^{(\pm)}_{N}(z,0)\rangle=\phi(z,0)|\Psi^{(\pm)}_{N}\rangle$ and $|\Phi^{(0)}_{N}(z,0)\rangle=\phi(z,0)|\Psi^{(0)}_{N}\rangle$
each of which obeys the one-dimensional time-dependent Schr\"{o}dinger equation
\begin{eqnarray}
i\hbar \frac{\partial}{\partial t}|\Phi^{(\pm)}_{N}(z,t)\rangle &=& \bigl (-\frac{\hbar^{2}}{2\mu}\frac{\partial^{2}}{\partial z^{2}} \pm 
\hbar g\sqrt{N}\bigr )|\Phi^{(\pm)}_{N}(z,t)\rangle, \nonumber \\
i\hbar \frac{\partial}{\partial t}|\Phi^{(0)}_{N}(z,t)\rangle &=&-\frac{\hbar^{2}}{2\mu}\frac{\partial^{2}}{\partial z^{2}}
|\Phi^{(0)}_{N}(z,t)\rangle, \label{TSE}
\end{eqnarray}
which show that we reduced the problem to a scattering process of a particle incident upon a potential 
$\pm \hbar g\sqrt{N}$, as depicted in Fig.~(\ref{fig-1}). Thus, the effect of the vacuum cavity mode is 
to induce a potential barrier and a well \cite{HW,USL} corresponding, respectively, to the eigenfunctions 
$|\Psi^{(\pm)}_{N}\rangle$ while the particles inside the cavity magnify considerable these potentials. 
Note that the external motion of atom experiences free evolution in the eigenstate $|\Psi^{(0)}_{N}\rangle$ 
for a z-independent cavity mode.

Supposing next that the peak of incident wave packet enters the cavity at time $t=0$ one arrives at the complete 
wave function of the atom-field system at time $t$ after the atom has left the interaction region:

\begin{eqnarray}
|\Phi(z,t)\rangle &=&\int d\chi A(\chi)e^{-i(\hbar\chi^{2}/2\mu)t}\bigl \{\frac{1}{\sqrt{2N}}\sum_{k \in \{+,-\}} \nonumber \\
&\times&[\rho^{(k)}_{N}e^{-i\chi z}\theta(-z)+\tau^{(k)}_{N}e^{i\chi z}\theta(z-L)]|\Psi^{(k)}_{N}\rangle \nonumber \\
&+& \sqrt{\frac{N-1}{N}}e^{i\chi z}\theta(z-L)|\Psi^{(0)}_{N}\rangle \bigr \} \nonumber \\
&=& \int d\chi A(\chi)e^{-i(\hbar\chi^{2}/2\mu)t}\sum_{s \in \{1,j,0\}}[R^{(s)}_{N}e^{-i\chi z} \nonumber \\
&\times& \theta(-z) + T^{(s)}_{N}e^{i\chi z}\theta(z-L)]|s\rangle, \label{FU}
\end{eqnarray}

where Heaviside's unit step function $\theta$ indicates on which side of the cavity the emitter can be found.
Here the reflection amplitudes of the particle incident upon a cavity while remaining in the states $|s\rangle$ 
$(s\in \{1,j,0\})$ are 

\begin{eqnarray}
R^{(1)}_{N} &=& [\rho^{(+)}_{N} + \rho^{(-)}_{N}]/(2N), \nonumber \\
R^{(j)}_{N} &=& \sqrt{N-1}R^{(1)}_{N}, \nonumber \\ 
R^{(0)}_{N} &=& [\rho^{(+)}_{N}- \rho^{(-)}_{N}]/\sqrt{4N}. \label{RA}
\end{eqnarray}

On the other hand the corresponding transmission amplitudes are given, respectively, by 

\begin{eqnarray}
T^{(1)}_{N} &=& [\tau^{(+)}_{N} + 2(N-1) + \tau^{(-)}_{N}]/(2N), \nonumber \\
T^{(j)}_{N} &=& \sqrt{N-1}[\tau^{(+)}_{N} - 2 + \tau^{(-)}_{N}]/(2N), \nonumber \\
T^{(0)}_{N} &=& [\tau^{(+)}_{N} - \tau^{(-)}_{N}]/\sqrt{4N}. \label{TA}
\end{eqnarray}
Other parameters are $\rho^{(\pm)}_{N}=i\alpha^{(\pm)}_{N}\sin[L\xi^{(\pm)}_{N}]\tau^{(\pm)}_{N}e^{i\chi L}$ 
and $\tau^{(\pm)}_{N}=e^{-i\chi L}\{\cos[L\xi^{(\pm)}_{N}]-i\beta^{(\pm)}_{N}\sin[L\xi^{(\pm)}_{N}]\}^{-1}$ 
with $\xi^{(\pm)}_{N}=\sqrt{\chi^{2} \mp \kappa^{2}\sqrt{N}}$, $\alpha^{(\pm)}_{N}=[\xi^{(\pm)}_{N}/\chi 
- \chi/\xi^{(\pm)}_{N}]/2$, and $\beta^{(\pm)}_{N}=[\xi^{(\pm)}_{N}/\chi + \chi/\xi^{(\pm)}_{N}]/2$, 
where $\kappa$ is the CM wave vector for which the kinetic energy $(\hbar \kappa)^{2}/2\mu$ equals the atom-vacuum
coupling energy $\hbar g$.  

The probabilities to find the atomic system in the state $|s\rangle$ is 
$P_{s} = |R^{(s)}_{N}|^{2} + |T^{(s)}_{N}|^{2}$. In particular, for fast atoms, i.e. 
$\chi^{2} \gg \kappa^{2}\sqrt{N}$, one obtains the results given in \cite{CuD}, i.e. 
$P_{f}=\bigl[N(1-\delta_{f1})-1+2\delta_{f1}\bigr]\bigl[N\delta_{f1} - 1 + \cos(gt\sqrt{N})\bigr]^{2}/N^{2}$ and
$P_{0}=\sin^{2}(gt\sqrt{N})/N$ with $f \in \{1,j\}$, while the interaction time is given by $t=\mu L/(\hbar\chi)$.

We focus further on the quantum tunneling of excited slow particles through multi-particle vacuum induced potentials. 
The transmission probability, i.e. the probability to find the excited atom on the right side of the cavity, can be 
calculated as $P^{(1)}_{T}=|T^{(1)}_{N}|^{2}$. A relatively simple expression for tunneling of the excited particle 
can be obtained if $\chi^{2} \ll \kappa^{2}\sqrt{N}$, namely

\begin{eqnarray}
P^{(1)}_{T}=\frac{1}{4N^{2}}\biggl [4(N-1)^{2} 
+ &\frac{1 + 4(N-1)\cos[L\kappa\sqrt[4]{N}]}{1 + (\kappa\sqrt[4]{N}/2\chi)^{2}\sin^{2}[L\kappa\sqrt[4]{N}]}\biggr], \label{TP}
\end{eqnarray}
since, for slow atoms, $\tau^{(+)}_{N} \to 0$ while $\rho^{(+)}_{N} \to -1$ which means that the dressed state component 
$|\Psi^{(+)}_{N}\rangle$ is always reflected. In particular, when $N=1$, i.e. a ultra - slow particle entering a empty 
cavity, one gets 
\begin{eqnarray*}
P^{(1)}_{T} = (4 + \kappa^{2}\sin^{2}{[\kappa L]}/\chi^{2})^{-1}, 
\end{eqnarray*}
that takes a maximum value if $\kappa L = \pi n$ where $n$ is an integer. On the other hand, for $N > 1$, maximum of 
$P^{(1)}_{T}$ occurs for $L\kappa\sqrt[4]{N}=2\pi n$, that is $P^{(1)}_{Tmax}=[1-1/(2N)]^{2}$, while minimum for 
$L\kappa\sqrt[4]{N}=(2n+1)\pi$, i.e. $P^{(1)}_{Tmin}=[1-3/(2N)]^{2}$. For moderate samples these expressions may help 
to determine the number of ground-state particles trapped inside the resonator within one emission wavelength. Interestingly, 
large atomic systems lead to a complete transfer of excited particles through the multiparticle-vacuum induced potentials 
because $P^{(1)}_{T} \to 1$ for $N \gg 1$ (see Eq. \ref{TP}). Additionally, if $N$ is large, $P^{(1)}_{T}$ is independent 
on the width of these potentials and on atomic velocities providing that the particle's kinetic energy $E$ is in the 
classically forbidden range (i.e. $E < \hbar g\sqrt{N}$). 
\begin{figure}[b]
\includegraphics[width=7.5cm]{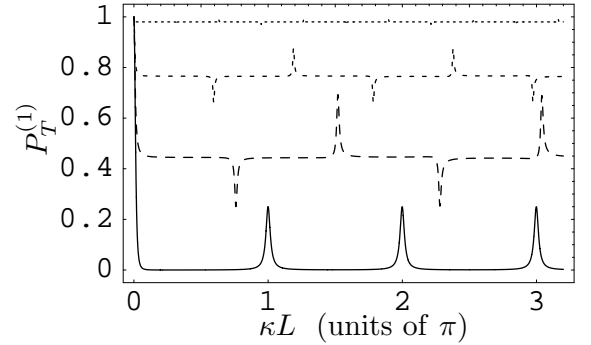}
\caption{\label{fig-2} Plot of the transmission probability $P^{(1)}_{T}$ against $\kappa L$. 
Here $\chi/\kappa=0.03$ while the number of atoms from the top to the bottom are, respectively, 
$N=100; 8; 3; 1$.}
\end{figure}

\begin{figure}[t]
\includegraphics[width=7.5cm]{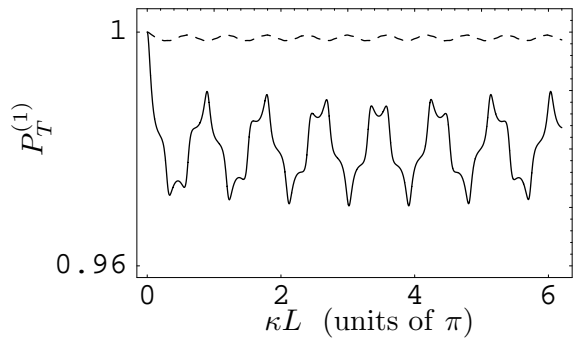}
\caption{\label{fig-3} The dependence of the transmission probability $P^{(1)}_{T}$ as function of $\kappa L$. 
The solid line is plotted for $N=100$ and $\chi/\kappa=1.01$ while the dashed curve is for $N=2000$ and $\chi/\kappa=5$, 
respectively.}
\end{figure}

%
Figures (\ref{fig-2}) and (\ref{fig-3}) depict, for different number of involved particles, the transmission probability 
$P^{(1)}_{T}$ as function of the external parameter $\kappa L$. As mentioned also above, by increasing the number of 
atoms inside the cavity, the tunneling probability $P^{(1)}_{T}$ approaches unity. This effect can be understood by 
inspecting the system's wave function (\ref{FU}) in the dressed-state picture. 
Remarkable, larger atomic samples, i.e. $N \gg 1$, allow to represent this state as follows

\begin{eqnarray}
|\Phi(z,t)\rangle=\int d\chi A(\chi)e^{-i(\hbar\chi^{2}/2\mu)t}e^{i\chi z}\theta(z-L)|\Psi^{(0)}_{N}\rangle. \label{FUA}
\end{eqnarray}

Thus, for many atoms inside the cavity, the contribution to the total wave function $|\Phi(z,t)\rangle$ resulting from the 
dressed state components $|\Psi^{(\pm)}_{N}\rangle$ are suppressed at resonance. This means that the whole system is into 
a quantum state characterized by the eigenfunction $|\Psi^{(0)}_{N}\rangle$ for which the slow atoms passing through the 
vacuum cavity mode experience free evolution. Moreover, $|\Psi^{(0)}_{N}\rangle$ represents a multiparticle dark state 
because the emitters that traverse through the resonator do not radiate photons into the cavity mode as the emission 
probability, i.e. $P_{0}=|R^{(0)}_{N}|^{2} + |T^{(0)}_{N}|^{2}$, tends to zero when $N \gg1$. Note that, in general, 
$|\Psi^{(0)}_{N}\rangle$ consists form a superposition of the first atom (which enters the cavity having $N-1$ 
indistinguishable ground-state particles) being excited and any other single atom excited from those that are inside the 
resonator while the first atom is leaving the vacuum cavity mode in its ground state. Analyzing then Eqs.~(\ref{RA}) we 
realize that all the reflection amplitudes (i.e. $R^{(1)}_{N}$, $R^{(j)}_{N}$ and $R^{(0)}_{N}$) vanish when $N \gg 1$. 
On the other hand, the transmission amplitude $T^{(1)}_{N} \to 1$ while $T^{(j)}_{N}$ and $T^{(0)}_{N}$ are zero for 
larger samples (see Eqs.~\ref{TA}). Thus, the quantum state of the moving atom is frozen when it passes through the 
cavity system because $|\Psi^{(0)}_{N}\rangle \approx |1\rangle$, if $N \gg 1$, while the transmission probability equals to 
unity in this particular case as was shown in Fig.~(\ref{fig-2}) and Fig.~(\ref{fig-3}) for various atomic velocities.

Finally, the peaks and the dips shown in Fig.~(\ref{fig-2}) are due to the induced photon emission into the cavity mode when 
the emitters traverse the interaction region \cite{HW,USL}. These sharp resonances are very sensitive on the velocity spread 
$\delta v$ of particles and may dissapear when averaging over a small, but not tiny, velocity range as was shown in \cite{HW} 
for single atom systems, i.e. for $N=1$. However, the magnitude of these maxima and minima decreases as the number of atoms 
increases (see Fig.~\ref{fig-2}). 

Concluding, an excited slow atom tunnels freely through a vacuum cavity mode filled with $N-1$ motionless indistinguishable 
ground-state particles. The reason for this is the trapping of the atom (incident on the cavity mode) into its upper state 
when it moves through the cavity and the corresponding decoupling of this particular atom from the interaction with the 
environment such that the emitter does not {\it feel} the induced potentials. 

{\it Acknowledgments:} I would like to thank Professor C. H. Keitel for valuable discussions.

%
{\small $^\ast$ On leave from \it{Technical University of Moldova, Physics Department, 
\c{S}tefan Cel Mare Av. 168, MD-2004 Chi\c{s}in\u{a}u, Moldova.}}

\end{document}